\documentclass{ieeeaccess}
\usepackage{cite}
\usepackage{amsmath,amssymb,amsfonts}
\usepackage{algorithmic}
\usepackage{textcomp}
\def\BibTeX{{\rm B\kern-.05em{\sc i\kern-.025em b}\kern-.08em
    T\kern-.1667em\lower.7ex\hbox{E}\kern-.125emX}}

\usepackage{graphicx}
\usepackage{latexsym}
\usepackage{amssymb}
\usepackage{amsmath}
\usepackage{url}
\usepackage{comment}
\usepackage[tight]{subfigure}
 \usepackage{multirow}
\usepackage{booktabs}
\usepackage{placeins}

\usepackage[usenames,dvipsnames]{xcolor}

\begin{document}
\sloppy
\history{Date of publication xxxx 00, 0000, date of current version xxxx 00, 0000.}
\doi{10.1109/ACCESS.2017.DOI}

\title{Towards robust audio spoofing detection: a detailed comparison of traditional and learned features. }

\history{Date of publication xxxx 00, 0000, date of current version xxxx 00, 0000.}
\doi{10.1109/ACCESS.2017.DOI}

\author{\uppercase{Balamurali B T}\authorrefmark{1},
\uppercase{Kin Wah Edward Lin\authorrefmark{1}}, \uppercase{Simon Lui}\authorrefmark{2}, \IEEEmembership{Member, IEEE}, \uppercase{Jer-Ming Chen\authorrefmark{3}}, and \uppercase{Dorien Herremans}\authorrefmark{1,4}, 
\IEEEmembership{Senior Member, IEEE}}

\address[1]{Information Systems, Technology, and Design Pillar, Singapore University of Technology and Design, Singapore }
\address[2]{Tencent Music Entertainment, Shenzhen, China }
\address[3]{Science Pillar, Singapore University of Technology and Design, Singapore }

\address[4]{Institute of High Performance Computing, A*STAR, Singapore}
\tfootnote{This research is supported by the Education Research Funding Programme, National Institute of Education (NIE), Nanyang Technological University, Singapore, under grant number AFD 05/15 SL. 
The views expressed in this paper are the authors' and do not necessarily represent the views of the host institution. Further, this work is also partly supported by SUTD SRG ISTD 2017 129.
}

\title{Towards robust audio spoofing detection: a detailed comparison of traditional and learned features}

%
%
%


\markboth
{B T \headeretal: Towards robust audio spoofing detection: a detailed comparison of traditional and learned features}
{B T \headeretal: Towards robust audio spoofing detection: a detailed comparison of traditional and learned features}

\corresp{Corresponding author: Balamurali B T (e-mail: balamurali\_bt@sutd.edu.sg).}



\begin{abstract}
Automatic speaker verification, like every other biometric system, is vulnerable to spoofing attacks. Using only a few minutes of recorded voice of a genuine client of a speaker verification system, attackers can develop a variety of spoofing attacks that might trick such systems. Detecting these attacks using the audio cues present in the recordings is an important challenge. Most existing spoofing detection systems depend on knowing the used spoofing technique. With this research, we aim at overcoming this limitation, by examining robust audio features, both traditional and those learned through an autoencoder, that are generalizable to different types of replay spoofing. Furthermore, we provide a detailed account of all the steps necessary in setting up state-of-the-art audio feature detection, pre-, and postprocessing, such that the (non-audio expert) machine learning researcher can implement such systems. 
Finally, we evaluate the performance of our robust replay spoofing detection system with a wide variety and different combinations of both extracted and machine learned audio features on the `out in the wild' ASVspoof 2017 dataset. This dataset contains a variety of new replay spoofing configurations. Since our focus is on examining which features will ensure robustness, we base our system on a traditional Gaussian Mixture Model-Universal Background Model (GMM-UBM). We then systematically investigate the relative contribution of each feature set. The fused models, based on both the known audio features and the machine learned features respectively, have a comparable performance with an Equal Error Rate (EER) of ~12. The final best performing model, which obtains an EER of ~10.8, is a hybrid system that contains both known and machine learned features, and is trained on an augmented dataset, thus revealing the importance of incorporating both types of features when developing a robust spoofing prediction model. 
\end{abstract}

\begin{keywords}
Audio Classification, Audio Spoofing, Autoencoders, Countermeasures, Replay Attacks, GMM-UBM. 
\end{keywords}

\titlepgskip=-15pt

\maketitle



%



\section{Introduction}

\PARstart{S}{ince} the dawn of Hacker Culture in the 50s and 60s~\cite{levy1984hackers}, enthusiasts have challenged themselves to overcome the limitations of software systems in order to achieve clever and creative outcomes~\cite{galston2004internet}. Security hackers in particular, have posed a threat by breaching defenses and exploiting the weaknesses in networks and systems. Biometric authentication systems such as automatic speaker verification (ASV) are not exempt from this threat. For instance, in audio spoofing attacks, fraudsters alter an audio recording of a voice, such that it mimics a target speaker's voice to access a system protected by speaker verification~\cite{1,2}. In replay attacks, imposters present speech samples recorded from a genuine client to a verification system~\cite{3,4,5}. 

Given the recent advances in audio processing technology, it is becoming easier to synthesize speech such that it sounds like a given target speaker. These technologies can be used by security hackers to break into ASV systems~\cite{6,7}. 
In addition to synthesizing speech, one could also use voice conversion methods that enable the conversion of utterances of one speaker to make them sound as if spoken by another speaker~\cite{8,9,10}. Given these advances, it is important to investigate whether we can discriminate original speech from spoofed speech recordings, which is the problem we tackle in this research. More specifically, our research contributes to a robust system for audio spoofing detection without knowing which spoofing technique is used in an attack. 
This challenge gets even more complicated if the discrimination has to be done only using only the audio data, without any other meta data~\cite{11}.

Existing countermeasures that try to detect specific spoofing attacks typically make use of prior knowledge about the used spoofing algorithm~\cite{12,13,14,15}. As a result, these countermeasures are not generalizable to varying spoofing attacks~\cite{16}.  A countermeasure for audio spoofing detection typically consists of two parts (see Figure~\ref{fig:architecture}). A first part dealing with features extraction and pre/postprocessing of the audio signal, and a second part consisting of a model that determines whether the audio is genuine or spoofed. During the system development, the spoofing prediction accuracy of the model is often compared for different audio features, so as to reach the highest accuracy~\cite{17,18}. We make this comparison explicit in our experiment section.

\Figure[ht](topskip=0pt, botskip=0pt, midskip=0pt){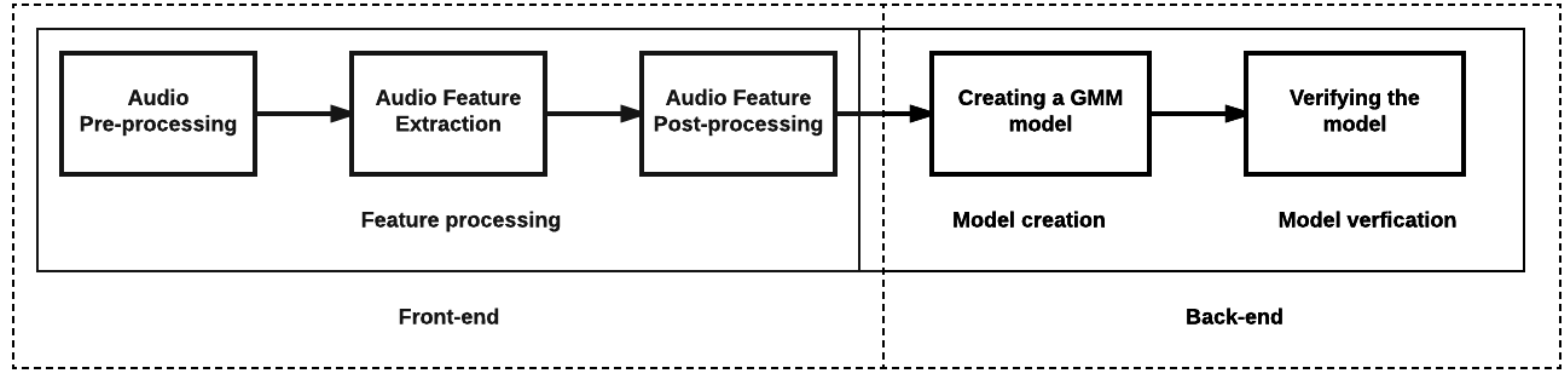}
{Architecture of proposed spoofing detection system. \label{fig:architecture}}

Developing spoofing prediction systems that are generalizable to a varying range of spoofing techniques, hinges on incorporating audio features that are robust, such that they require little recalibration to detect novel spoofing attacks~\cite{11, yu2017dnn}. One way to achieve this robustness is by investigating the spoofing detection performance when using various audio features. This is the key contribution of this manuscript. We leverage the dataset from the Second Automatic Speaker Verification Spoofing and Countermeasures Challenge, ASVspoof 2017~\cite{19}, and investigate the performance a wide variety of audio features on different types of replay spoofing attacks. In addition using a variety of traditional audio features, we train an autoencoder and use it to augment the training dataset. 

The second step in developing a spoofing countermeasure is to use the extracted audio features to train a spoofing prediction model. The models typically used in speech/speaker related applications have evolved a lot over the past 40 years. In the past, researchers often used systems based on Discrete Vector Quantization (VQ)~\cite{20}. The state-of-the-art then moved to Gaussian Mixture Model (GMM) solutions~\cite{21}, and more recently into factor analysis based on $i$-vector frameworks~\cite{22}. In this research, we explore the effect of different audio features in a GMM system. 

The ASVspoof 2017 database was designed mainly to assess the detection accuracy of replay spoofing attacks especially for `out in the wild' conditions, meaning without knowing the exact spoofing technique configuration used in the replay attack. In order to do so, the majority of testing data included in this database originates from different, unseen configurations of the spoofing algorithms compared to the training and development set~\cite{11}. Such `out in the wild' conditions often necessitate a new audio feature space to accommodate for different replay spoofing techniques, and typically results in inferior performance when evaluating the models. This again confirms the need for a generalized countermeasure. Such a countermeasure could be achieved either by identifying robust audio features, or by improving an automatically learned feature space of spoofed recordings through an autoencoder. In this research, we explore both approaches in depth. 


The remainder of this paper is organized as follows. The developed spoofing detection system is described in Section~\ref{sec:system} and III. These sections include an overview of the audio features used in this investigation and description of the automatic feature extraction process. We provide a very thorough description of, not only how to extract and learn audio features, but also which pre- and postprocessing steps are required for obtaining the best results, thus providing guidance for the audio laymen to tackle this type of spoofing challenge. In addition, this section describes the implemented autoencoder, and how it can be used to augment the dataset. Details about the experimental setup can be found in Section~\ref{sec:setup}, which describes the dataset used in this investigation. The next section describes the results (Section~\ref{sec:results}), in which we discuss the most robust features. Finally, Section~\ref{sec:conclusion} contains our conclusions and final remarks.

\section{A robust hybrid replay spoofing detection architecture}
\label{sec:system}

The hybrid spoofing detection system developed in this paper consists of two branches that each have a unique way of processing the input audio files (see Figure~\ref{fig:architecture}). Large sets of selected audio features are first extracted from the preprocessed audio files. These features are less redundant and more compact than the original audio signal. In the first arm of the algorithm, the audio features are postprocessed and passed along to the prediction module. In the second arm, however, they are first fed to an autoencoder before postprocessing. For both arms of the algorithm, a GMM-UBM model is built (see Section~\ref{sec:backend}) that can predict the authenticity of a given audio file. Finally, both model outputs are fed to a fusion model, which will calculate a hybrid estimation of authenticity. 

In what follows, we describe the different modules of our spoofing detection system in more detail, followed by an in-depth account of the extracted audio features. 


\subsection{Feature Processing}
\begin{figure*}[t]
\centering
\includegraphics[width=.94\textwidth]{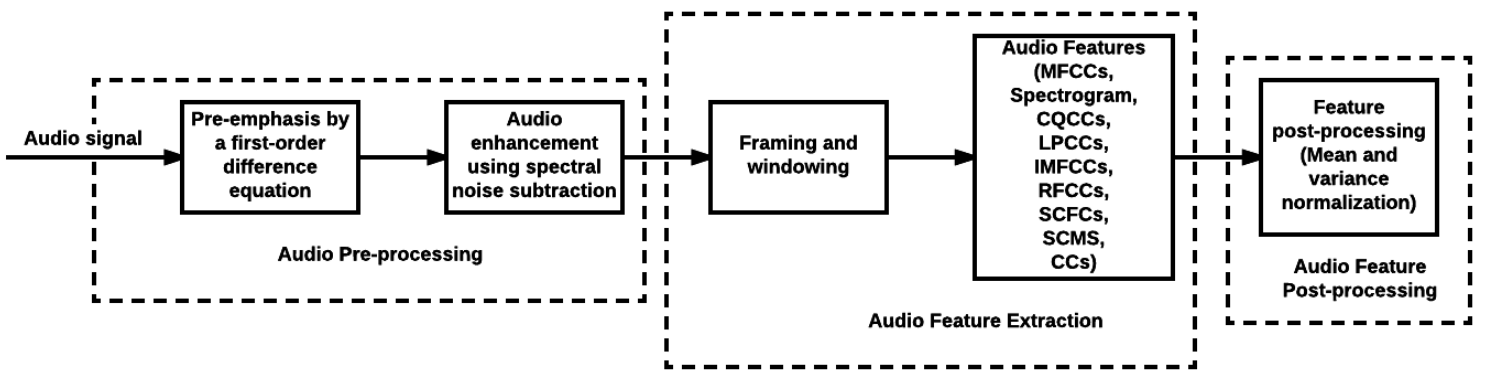}
\caption{Architecture of the feature processing unit, which includes audio preprocessing, audio feature extraction, and audio feature postprocessing.}
\label{fig:audio}
\end{figure*}

The first part of our system relates to audio feature processing and consists of a number of steps, as described below. Figure~\ref{fig:audio} zooms in on the audio processing module in arm 1, without autoencoder.

\subsubsection{Audio preprocessing}
The audio preprocessing module consists of two parts: pre-emphasizing and enhancing the audio file. The audio recordings present in the corpus are first pre-emphasized by applying a first order difference equation~\cite{25}. The main goal of pre-emphasis is to boost the amount of energy present in the high frequencies. The recordings are further enhanced by removing the noise using a spectral noise subtraction method whereby an estimate of the noise spectrum is subtracted from the speech power spectrum and the negative differences are set to zero. This new power spectrum is then recombined with the original phase of the audio signal so as to reconstruct an enhanced version of the time waveform~\cite{26,27}. 
This preprocessing module is implemented using the `Voicebox tool box for speech processing' in Matlab~\cite{28}.

\subsubsection{Audio feature extraction }
The enhanced audio signals are divided into frames, so as to form segments which are more stationary. Next, a hamming window is applied, with 50\% overlap. A large number of audio features, 11 feature sets in total, are then extracted from every frame. These feature sets include mel-frequency cepstral coefficients (MFCCs), spectrogram, constant Q cepstral coefficients (CQCCs), linear predictive cepstral coefficients (LPCCs), inverted mel-frequency cepstral coefficients (IMFCCs), rectangular filter cepstral coefficients (RFCCs), linear filter cepstral coefficients (LFCCs), sub-band centroid frequency coefficients (SCFCs), sub-band centroid magnitude coefficients (SCMCs), and complex cepstral coefficients (CCs). 
A more detailed explanation of these features can be found in Section~\ref{sec:features}.

\subsubsection{Autoencoder}

\begin{figure*}[h]
\centering
\includegraphics[width=.94\textwidth]{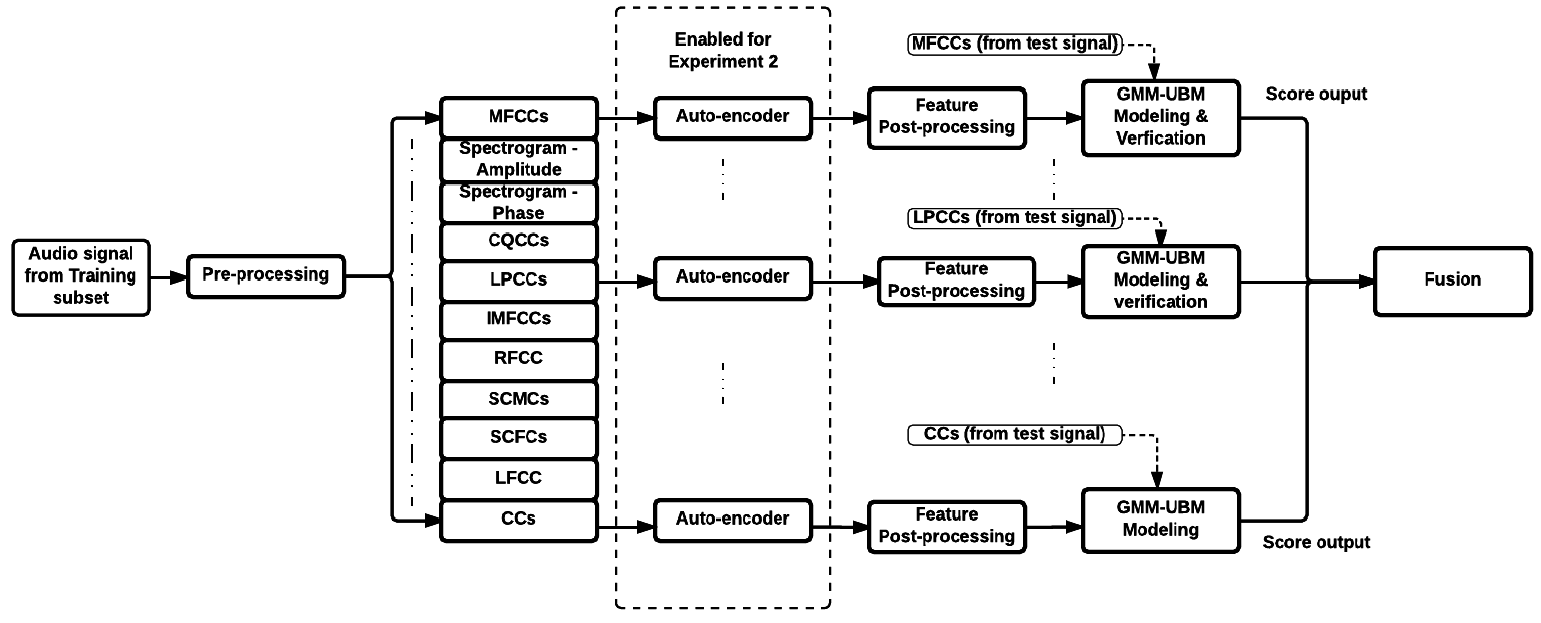}
\caption{The autoencoder arm of the proposed robust hybrid spoofing detection system. }
\label{fig:arm2}
\end{figure*}

In the second arm of the system, we insert a pretrained autoencoder before feature postprocessing, as is displayed in Figure~\ref{fig:arm2}. An autoencoder is a special type of feedforward neural network with fully connected layers that is trained by matching its input to its output~\cite{23,24}. The encoder compresses the input into a lower-dimensional space (also called `code'), which can then be used by the decoder to reconstruct the output. After training, the output of an autoencoder will typically not be exactly the same as the original input, but it will be a closely resembling, slightly degraded version of the input~\cite{23,24}. 
In addition to learning new features (i.e., the code), the autoencoder is also used to augment the dataset, as explained in Section~\ref{sec:aug}. 

\subsubsection{Audio feature postprocessing}
In order to further reduce any distortion of the data caused by noise contamination, the original audio features (arm 1) or the encoder features (arm 2) are mean and variance normalized. This is achieved by linearly transforming the features such that they have the same statistics within the segment. It was reported that such postprocessing can significantly improve the performance of speaker/speech recognition systems~\cite{29,30}. Figure~\ref{fig:audio} outlines the entire audio preprocessing and feature extraction process in arm 1.

\vskip 1.5cm

\subsection{Feature modelling and prediction}
\label{sec:backend}

\subsubsection{GMM-UBM}

The second module of the system performs the actual spoofing detection. We use a GMM-UBM for this investigation, which is created for each of the 11 feature sets in a two-stage process. Firstly, a universal background model (UBM) is created using the entire training set for a given feature set. As this pool includes both genuine and spoofed data, the resulting background model captures the feature space of both types of recordings. UBM is a Gaussian Mixture Model created using a binary splitting expectation-maximization (EM) procedure~\cite{21,31}. In the second stage, two GMM models, one for genuine and one for spoofed recordings, will be created by adapting the UBM using maximum a posteriori (MAP) estimation to fit the respective data. Finally, in the prediction phase, the  likelihood of the audio features extracted from test segments can be calculated based on each of the two models. The ratio of both of these likelihood probabilities is indicative of the predicted class. 

\begin{figure*}[t]
\centering
\includegraphics[width=.94\textwidth]{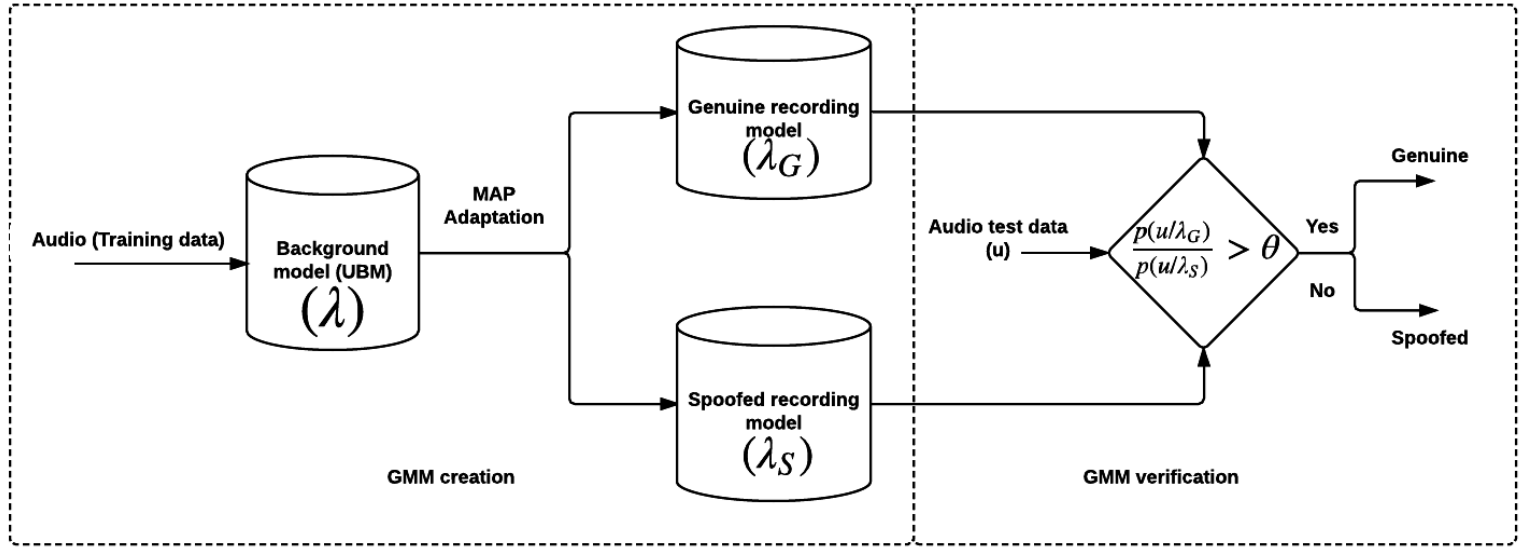}
\caption{Spoofing prediction using a GMM-UBM.}
\label{Figure3}
\end{figure*}

The process of creating the initial GMM models, one for each feature set, and predicting the class for a new instance is shown in Figure~\ref{Figure3}. In a first step, the postprocessed audio features from the training set are used to create a universal background model ($\lambda$). From this $\lambda$, a genuine recording model $(\lambda_{G})$ and a spoofed recording model $(\lambda_{S})$  are created using the maximum a posteriori (MAP)  adaptation procedure with genuine and spoofed audio features, respectively. These $\lambda_{G}$ and $\lambda_{S}$ can then be used to calculate the log-likelihood of a test audio file $u$ (see Figure \ref{Figure3}). For $\lambda_{G}$ this results in $p(u/\lambda_{G})$) and for $\lambda_{S}$ the probability is $p(u/\lambda_{S})$. The ratio of both of these probabilities, $\frac{p(u/\lambda_{G})}{p(u/\lambda_{S})}$ can be used to predict the class.

A number of parameters need to be set when creating UBM models. These include the number of Gaussian components and the number of expectation maximization (EM) iterations in the final binary split. Even though the number of Gaussian components required to model various audio features is different, we decided to set this number to 64, based on our investigation of audio samples from the development set. The number of EM iterations was set to 30.

\subsubsection{Fusion model}

Given that we create $22 (11 + 11)$ individual GMM-UBM models, $11$ for each feature set in arm 1 and $11$ for each feature set after the autoencoder in arm 2, a fusion model is needed to merge the individual prediction results. We included a logistic regression fusion procedure~\cite{32,33} to make a final prediction. 

For this procedure, we determine the fusion parameters (i.e., weights for the likelihood of each separate model, and a shifting factor), using the development set. These fusion parameters are then used to combine multiple likelihoods, one corresponding to the model for each feature set, to produce a final likelihood, which will determine if the audio file is spoofed or genuine.

\section{Known versus Learned Audio Features }
\label{sec:features}

The extracted audio features, together with the learned encoded representation from the autoencoder are discussed below. Finally, we describe how the autoencoder was used to augment the dataset and create a new set of instances for training. 


\subsection{Existing audio features}
\label{sec:audio}
A total of 11 sets of known features are extracted from the dataset. Below we discuss each of these feature sets.

\subsubsection{Mel-frequency cepstral coefficients (MFCCs)}

The first set of features, consisting of MFCCs, focuses on the perceptually relevant aspects of the speech spectrum. They are arguably the most commonly used speech features in the speech/speaker recognition arena~\cite{34,35}. The extraction process of MFCCs can be summarized as follows. First, the Discrete Fourier transform (DFT) of the input audio frame is taken to obtain the audio spectrum. This spectrum will  indicate the amount of energy present in the various frequency bands. As the human hearing is very sensitive to lower frequencies, but less able to distinguish adjacent high frequency sounds, it was chosen to use a representation that accounts for this: mel-spectrum. To calculate this representation, the spectrum is warped into a mel scale, by using a bank of triangular filters (i.e., a set of overlapped non-linear mel-filter banks whose bandwidth gets narrow at low frequencies and wider at higher frequencies). 
Once the spectrum is warped, the logarithm of the energy present in the various regions of the speech spectrum is estimated. Finally, this log spectrum is transformed back to the time domain, thus resulting in MFCCs~\cite{36}.

In the speech recognition arena, it is common to use the first 12-14 MFCCs extracted from a stationary speech frame along with their deltas (first derivatives) and delta-deltas. The first and second derivatives of the MFCCs carry information about speech dynamics.
In this paper, we used 24 mel-filters and extracted 13 MFCCs together with the energy of  every frame (i.e., the zeroth MFCC corresponds to energy present in the frame). By including the deltas and delta-deltas (i.e., the first and second order derivatives of MFCCs) there will be a total of 42 features per frame (i.e., 14 MFCCs, 14 deltas and 14 delta-deltas).
Mathematically, MFCC feature extraction can be summarized as follows~\cite{36,37}:

\begin{equation}
\text{MFCC}(q)=\sum_{m=1}^{M}\log \left[ \text{MF}(m) \right] \cos \left\lbrace \frac{q(m-0.5)\pi}{M}  \right\rbrace\\
\end{equation}
\begin{equation}
\text{MF}(m)=\sum_{k=1}^{K} \left| X_{\text{DFT}}(k)  \right|^{2}H_{m}(k)
\end{equation}

whereby $H_m (k)$ is the $m^{th}$ mel-filter bank, $\text{MF}(m)$ is the mel-frequency spectrum, $M$ is the number of mel-filter banks, $q$ the number of MFCCs, $k$ is the DFT index and $K$ is the total number of DFT indices, $X_{\text{DFT}}$ is the DFT of input audio frame.\par

\vspace{2pt}
\subsubsection{Other Ceptral Coefficients related to MFCCs}

In addition to MFCCs, we also included Rectangular Filter Cepstral Coefficients (RFCCs), Linear Filter Cepstral Coefficients (LFCCs) and Inverted Mel-frequency Cepstral Coefficients (IMFCCs). The extraction process of these feature sets is very similar to that of MFCCs. There are, however, subtle differences in the selection of filters and the chosen frequency scale~\cite{17}. For instance, RFCCs use a bank of 24 uniform non-overlapping rectangular filters distributed over a linear frequency scale. The procedure for calculating LFCCs is similar to that of RFCCs~\cite{38}, however, it uses triangular filters instead of rectangular filters~\cite{17,39}.
Finally, to calculate IMFCCs, overlapping triangular filters are linearly placed over an inverted-mel scale~\cite{40}. This means that IMFCCs will emphasize the higher frequency region of the spectrum, i.e., opposite to human perception. Similar to MFCCs, we have also used the deltas and delta-deltas of these audio features in our experiments.

\subsubsection{Spectrogram}

    Spectrograms provide a visual representation of the spectrum of frequencies present in an audio signal over time. They are nowadays widely used as features in image/audio classification and separation systems~\cite{41, svs18}. When tested on a sound event or speech classification task, they have shown to provide a significant improvement in classification performance when supplemented with other traditional audio features such as MFCCs and LPCCs ~\cite{42,43}. In this paper, we obtain the spectrum of an audio signal by calculating the short-time Fourier transform (STFT) of the audio input. This can then be unwrapped to get the amplitude and phase information~\cite{16}. Both of these are considered as separate features for this investigation.

\subsubsection{Constant $Q$ Cepstral Coefficients (CQCCs)}

CQCCs make use of a perceptually motivated time-frequency analysis known as constant $Q$ transform (CQT). 
The frequency bins in Fourier-based approaches are often regular spaced and will result in a variable $Q$ factor $\left( Q=\frac{center frequency of band}{bandwidth}  \right)$ \cite{16,44,45}, based on the center frequency of each particular band. 
In CQT, however, the bins are spaced geometrically, in order to ensure a constant Q factor.  In stark contrast to the Fourier approach, CQT offers a higher frequency resolution at lower frequencies and a higher temporal resolution at higher frequencies. The extraction process of CQCCs begins by taking the constant-$Q$-transform of the input audio frame. The power spectrum is then computed and its logarithm is calculated.  Since the $k$ bins in a constant $Q$ transforms are each on a different scale, the log power spectrum has to be resampled before applying the discrete cosine transform (DCT). This resampling is achieved by converting the geometric space to linear space. This involves a down-sampling operation over the first $k$ bins (i.e., the low frequency part) and an up-sampling operation for the remaining bins (i.e., the high frequency). This linearization of the frequency scale of the CQT further preserves the orthogonality of the DCT outputs~\cite{44}. In this investigation, 20 CQCCs, their deltas and delta-deltas have been extracted from each audio frame, thus forming a total of 60 features.


\subsubsection{Linear predictive cepstral coefficients (LPCCs)}

Another standard set of features widely used in speech recognition are LPCCs~\cite{36,46}. They are computed from the smoothed auto-regressive power spectrum of an audio frame. LPCC extraction begins with the estimation of the linear predictive coding (LPC) coefficients of an audio frame. These linear predictive coefficients are converted to LPCCs by using a recursion algorithm as shown below~\cite{47}. In this investigation, 16 LPCCs have been extracted from every audio frame.\par

Consider the speech input $s(n)$ to an all-pole LPC filter, which has $p$ linear predictive coefficients, and a prediction error signal of  $e(n)$. Let $E_e$ be the power of this error signal. Now the $p$ coefficients $[a_0,a_1,\dots,a_{p-1}]$ can be recursively converted to $n$ LPCCs $[L_0,L_1,\dots,L_{n-1}]$ as follows for each LPCC. 




\noindent For  $m = 0$:
\begin{equation}
L_{m}=\ln (E_{e})\\
\end{equation}

\noindent For  $1\leq m\leq p$:
\begin{equation}
L_{m}=-a_{m}+\frac{1}{m}\sum_{k=1}^{m-1}\left\lbrace -(m-k)a_{k}L_{m-k}  \right\rbrace  
\end{equation}

\noindent For  $p<m<n$:
\begin{equation}
L_{m} = \frac{1}{m}\sum_{k=1}^{p} \left\lbrace -\frac{(m-k)}{m}a_{k}L_{m-k}\right\rbrace 
\end{equation}

\vspace{12pt}

\subsubsection{Spectral sub-band Centroid Coefficients}


Spectral sub-band centroids, as the name suggest, represent the centroids of selected sub-bands of the spectrum. Both spectral sub-band centroid magnitude (SCM) and spectral sub-band centroid frequency (SCF) can be extracted from a given sub-band, whereby the characteristics of the latter one are similar to that of formant frequency~\cite{17,48}. 
Sub-band centroid features, supplemented with cepstral features, have shown to work very well for speech recognition~\cite{49}. 

The SCMC and SCFC extraction process starts by creating the spectrum for a given speech frame and dividing this into $k$ sub-bands, whereby each sub-band is defined by a fixed lower edge and an upper edge frequency. SCFC is then calculated from the average frequency for each sub-band, and weighted by the normalized energy of each frequency component in that sub-band. Similarly, SCMC is calculated as the average magnitude of each sub-band, weighted by the frequency of each magnitude component in that sub-band. For calculating the SCMC, we then take the logarithm of this result and apply DCT~\cite{17,48}. For our experiments, we have used 8,192 bins (sub-bands) of spectrum extracted from every 20ms audio frame.

\subsubsection{Complex Cepstral cofficients (CCCs)}

Cepstral analysis, often referred to as homomorphic filtering~\cite{50}, can be used to separate out various components in a speech production (i.e., source-filter) model. The Complex Cepstrum is defined as the inverse Fourier transform (IFT) of the logarithm of the Fast Fourier Transform (FFT) of a signal~\cite{51}. By taking the FFT of the speech signal, the convolution between the source and filter components in the time domain will be converted into their product in the frequency domain. The logarithm operator then transforms this product operation into a sum operation of both components. Finally, an IFT is taken to bring these summed components back into the time domain (quefrency domain)~\cite{52}.

The resultant complex cepstral coefficients (CCs) characterize the slow and fast varying components of speech. Slow-varying components (e.g. contribution of pitch) are concentrated in the upper part of the cepstral domain, whereas the fast-varying components (e.g. contribution of the vocal tract filter) are concentrated in the lower part~\cite{51}. Since CCs carry more speech specific information than most of the other cepstral coefficients that we discussed above, they typically result in higher speaker recognition performance~\cite{53}. In the experiments below, we have used the lower 50 cepstral coefficients extracted from every frame.

\subsection{Learned feature representation}

An autoencoder is trained for each feature set, such that a dense presentation of all original audio features is learned. Each of these resulting dense representations (one per feature set) are then concatenated to form the input for the final classification model.

To be able to train the autoencoders, a number of hyper-parameters are set, including encoding-size (code-size $ = 100$), number of layers in encoder and subsequently the decoder  (one layer for each), number of nodes per layer (ten per layer), and loss-function~\cite{58,59}. Figure \ref{Figure6} shows the autoencoder architecture used in our experiments for a particular feature vector of size $1\times n$.

\Figure[h]()[width=0.48\textwidth]{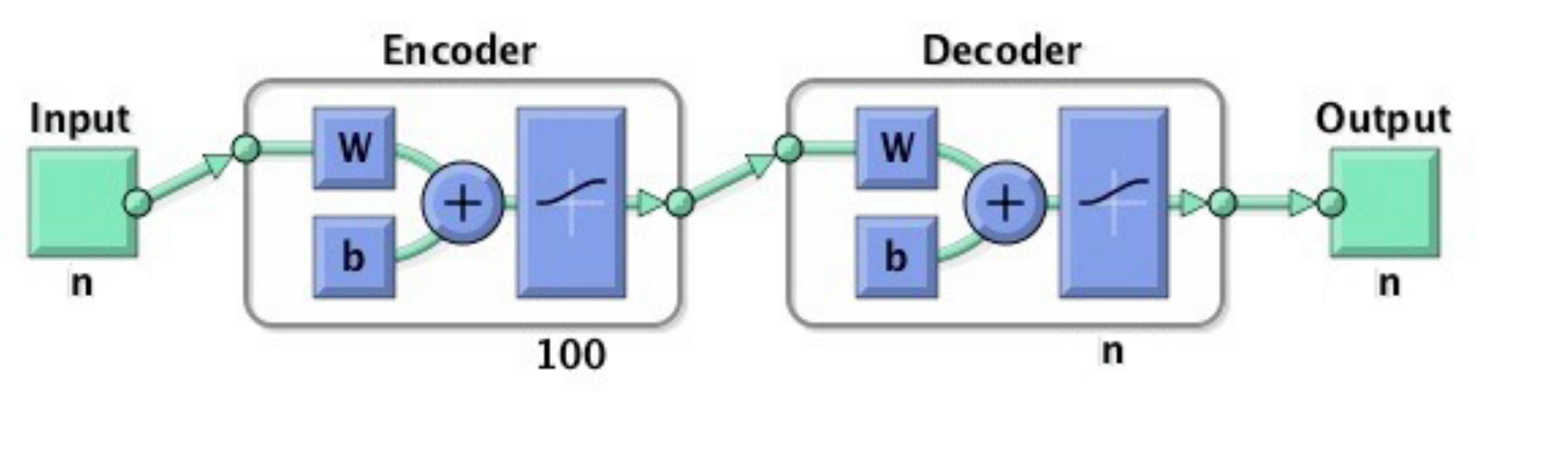}
{Architecture of the autoencoder used in this investigation, for a feature vector of size $n$. \label{Figure6}}

In the figure, the size of the input feature vector is $1\times n$. The code-size is set to 100, and the encoder/decoder each consist of 1 layer. 
Mean squared error (MSE) is used as the loss function. MSE is the recommended loss function for cases in which the input values are in the range of zero to one~\cite{24}. In this example, the encoder compresses the original $1\times n$ input to a lower dimensional code (in this case of size 100) and the decoder reconstructs the output of size $1\times n$ using this code. The encoder will be trained until it reaches the performance target. In our case, this in turn corresponds to achieving the lowest gradient, which corresponds to the lowest loss in prediction error. To finish the training in a finite time, the maximum number of epochs in the training phase is set to 200. This training cutoff was set after careful investigation of the evolution of the MSE loss when using different training samples, corresponding to each of the feature vectors. We found that the MSE is very low for 200 epochs and did not change too much in subsequent epochs.


\subsection{Data augmentation}
\label{sec:aug}

\Figure[h!]()[width=0.96\textwidth]{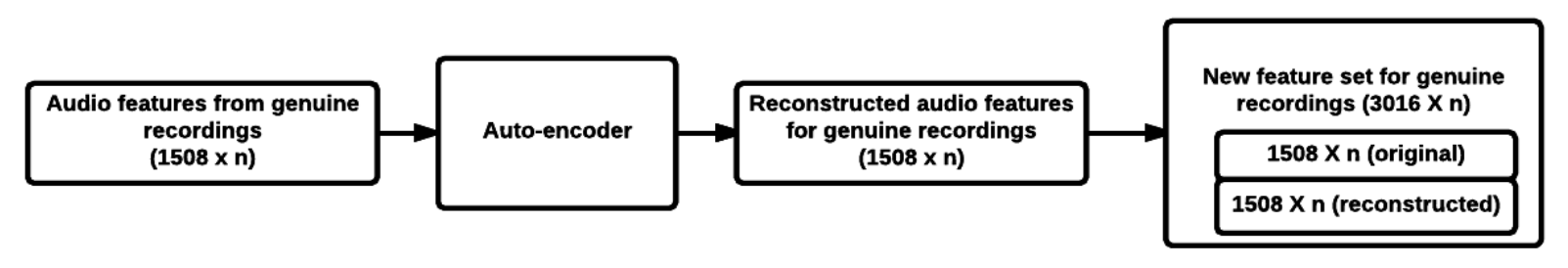}{Data augmentation (with $n$ number of features, and $1,508$ original genuine recordings). A similar procedure is performed on the spoofed recordings. \label{Figure7}}

In addition to creating a new, dense representation, the autoencoder trained on genuine recordings is also used to augment our dataset (see Figure~\ref{Figure7}). This is achieved by using the autoencoder to reconstruct the feature sets for the $1,508$ genuine recordings in the original training set. These $1,508$ reconstructions are then added to the training set. A similar procedure is performed on the spoofed recordings.


\section{Experimental setup}
\label{sec:setup}

We investigate the influence of both the known and learned feature representations on spoofing detection accuracy. 
The setup of our experiment, including dataset, evaluation methodology, are discussed below.


\subsection{Audio dataset - ASVspoof 2017}

\begin{table*}[h!]
\centering
\caption{Statistics of ASVspoof 2017 database.}
\label{Table1}
\begin{tabular}{cccccc}
\toprule
\multirow{2}{*}{Subset} & \multirow{2}{*}{Number of speakers} & Number of replay & Number of replay  & Number of non-replay  & Number of replay \\
~ & ~ & sessions & configuration & utterances & utterances\\
\toprule
Training & 10 & 6 & 3 & 1,508 & 1,508\\
Development & 8 & 10 & 10 & 760 & 950\\
Testing & 24 & 163 & 112 & 1,298 & 12,008\\
\bottomrule
\end{tabular}
\end{table*}

The `out in the wild' ASVspoof 2017 dataset (protocol V2) is used for our experiments~\cite{54}. This corpus originated from RedDots corpus3 and contains recordings collected by researchers using Android mobile phones~\cite{55,56}. The recordings include both replayed (spoofed) and non-replayed (original) utterances. The former (i.e., replayed) are captured versions of the original RedDots recordings, meaning that an utterance of an original target speaker was replayed through transducers of varying quality, and recorded using a mobile phone), whereas the latter (i.e., non-replayed utterances) are original recordings. Replayed utterances can be used to model a `stolen voice'. 

We chose to work with this dataset, as it contains audio from different, unseen configurations of the spoofing algorithms in the testing set. The aim of this paper is to assess the validity of different audio features when building a robust spoofing detection model, which remains effective without knowing which spoofing technique was used.

The ASVspoof corpus is divided into three subsets: training, development and testing set. The training and development set are used to train and validate our spoofing detection system, whereas the testing set is used to test the performance. Since heterogeneity in the data is highly essential for developing reliable spoofing countermeasures, no two-same-speaker recordings are included in any of the three subsets. Further, the data collection sites are also chosen to be distinct for all the three subsets~\cite{11,54}. More information about corpus is shown in Table~\ref{Table1}. 

\subsection{Equal Error Rate (EER) as the evaluation metric}

The Equal Error Rate (EER) is used as the primary metric to evaluate the performance of our spoofing detection system~\cite{11,57}. The false positive rate $(p_{fp}(\theta))$ and false negative rate $(p_{fn}(\theta))$ at a particular threshold are required to calculate EER: 

\begin{align}
p_{fp}(\theta)&=\frac{\text{Number of replay trials with likelihood} > \theta}{\text{Total number of replay trials}}\\
p_{fn}(\theta)&=\frac{\text{Number of non-replay trials with likelihood} \leq \theta}{\text{Total number of non-replay trials}}
\end{align}

The EER corresponds to the particular value of $\theta$ for which $p_{fp}(\theta)$ and $p_{fn}(\theta)$ are approximately equal. The lower the EER value, the better is the performance of the system. In this paper, the term `performance' refers to EER results, which can be seen as a measure of accuracy.

\subsection{Experimental methodology}

In order to thoroughly examine the effect of different audio features and the autoencoder, a total of three experiments are conducted, as represented in Figure~\ref{fig:arm2}. 

In \emph{Experiment 1}, the audio features explained in Section~\ref{sec:audio} are extracted from the training set and postprocessed to create individual feature-specific GMM models. This means that a UBM is created for each set of features (e.g. MFCCs, CQCCs, etc.), using the $1,508$ genuine and $1,508$ spoofed audio files in the training set. We evaluate and compare the performance of models based on each of these features separately. In addition, we explore a fused model that takes as input the likelihoods corresponding to the models of all of the feature sets investigated. 



In the \emph{second experiment}, an autoencoder is used to learn a new representation for each of the original feature sets, and to augment the training set. We again evaluate the performance of each feature set, and compare this to a hybrid system that is obtained by fusing the individual likelihoods obtained per feature set. 


The \emph{third experiment} is a fusion of all the results from Experiment 1 and 2, whereby the GMM-UBM likelihoods for each of the models based on individual features sets are combined using a logistic regression fusion.


\section{Results}
\label{sec:results}

We discuss the performance of the different features through three experiments as described below, followed by a comparison with other state-of-the-art models. 

\subsection{Experiment 1: Performance of known audio features}

The performance of models built on each of the different feature sets is displayed in Table~\ref{Table2}. The model based on \emph{CQCCs} performs best. This result aligns with previous studies conducted in this arena~\cite{37,44}.

Interestingly, the performance of the model based on \emph{MFCCs}, one of the de-facto features extracted in many of the speech/speaker recognition, is found to be the worst. As explained earlier, MFCCs capture perceptually relevant aspects of the speech spectrum. In a replay attack scenario, the imposter is playing speech samples captured from a genuine speaker, hence, it could be expected that the perceptual/ audible artifacts would be the same as that of the original audio, which explains the low performance when using MFCCs. 

Three models based on \emph{cepstral co-efficients} (RFCCs, LFCCs and IMFCCs), whose extraction process is very similar to that of MFCCs yet uses a different selection of filters and frequency scales, perform well. Among these three, the inverted-MFCCs, which models characteristics opposite to human perception, is the best performer. This falls in line with our previous hypothesis that replayed audio may have inaudible artifacts and that IMFCCs might be capturing cues outside of human perception.


\Figure[h!](topskip=0pt, botskip=0pt, midskip=0pt)[scale=0.8]{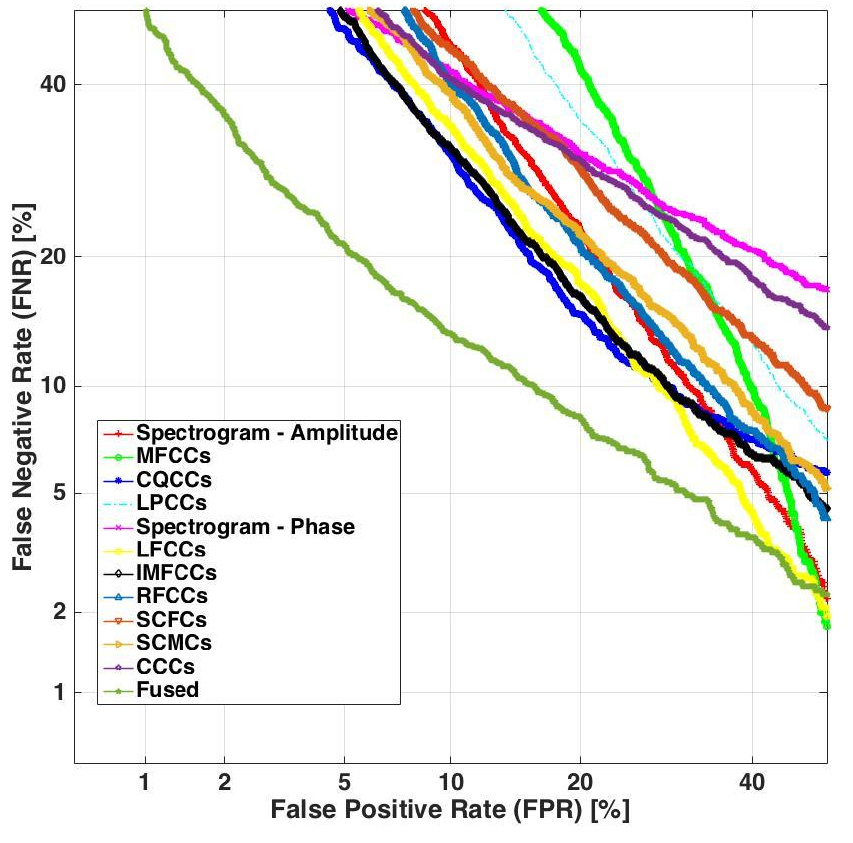}
{DET Curves showing the performance of spoofing the detection models for Experiment~1. \label{Figure8}}

\begin{table}[h!]
\centering
\caption{Results from Experiment 1.  }
\label{Table2}
\begin{tabular}{lc}
\toprule
\textbf{Features} & \textbf{EER}\\
\toprule
MFCCs & 27.2\\
CQCCs & \textbf{17.5}\\
Spectrogram - Amplitude &  20.9\\
Spectrogram - Phase & 26.8\\
LPCCs & 26.1\\
RFCCs & 20.3\\
LFCCs & 19.0\\
IMFCCs & \textbf{18.1}\\
SCMCs & 21.1\\
SCFCs & 24.0\\
CCCs & 25.9\\
Fused result & \textbf{12.2}\\
\bottomrule
\end{tabular}

\vspace{6pt}
\footnotesize \emph{The features resulting in the three best performing\\ models are indicated in bold.}
\end{table}

The model based on \emph{spectral amplitude} is a good candidate to distinguish spoofed recordings from genuine. A related model, using \emph{spectral phase} information, however, did not perform up to the expectations set in~\cite{16}.

The performance of models that use features based on \emph{spectral sub-band centroids} (e.g. SCMCs and SCFCs) is similar, with a slight edge for amplitude based features (i.e., SCMCs) versus frequency based features (i.e., SCFCs). This trend is also observed when comparing spectral amplitude versus spectral phase. The effect can be due to the way that an imposter tries to capture genuine speech samples, as the energy/amplitude of the captured speech will be directly related to the placement of the capturing device with respect to the mouth position of the genuine speaker. This could be reflected in the extracted features.



Finally, models built using \emph{complex cepstral coefficients and LPCCs} do not perform as to expected. The former are expected to carry more speech specific information. The latter inherit advantages of LPC (i.e., linear prediction coefficients), which are related to the speech production (source-filter) model. Both of these features result in models with a similar performance to those using MFCCs.

In order to achieve better spoofing detection performance and to overcome the limitation in performance of individual features, the results from models based on different feature sets have been merged using logistic regression fusion. As expected, the resulting hyrbid model has a far better overall performance.

The results of Experiment 1 are visualised in Figure~\ref{Figure8} by using an detection error trade-off (DET) graph. DET graphs show the performance of detection tasks that involve a trade-off error. In this case, false negative and false positive rates for each set of features 
 are plotted against each other in the DET graph. The EER can be observed from the DET graph as the point where the percentage of false negatives equals the false positive rate~\cite{60}. Better performance is reflected by closer proximity to the origin. This graph confirms our previous conclusions and identifies the fused model as the best performing.


\subsection{Experiment 2: Performance with autoencoder features and data augmentation}

\begin{figure}[h!]
\centering
\includegraphics[scale=0.74]{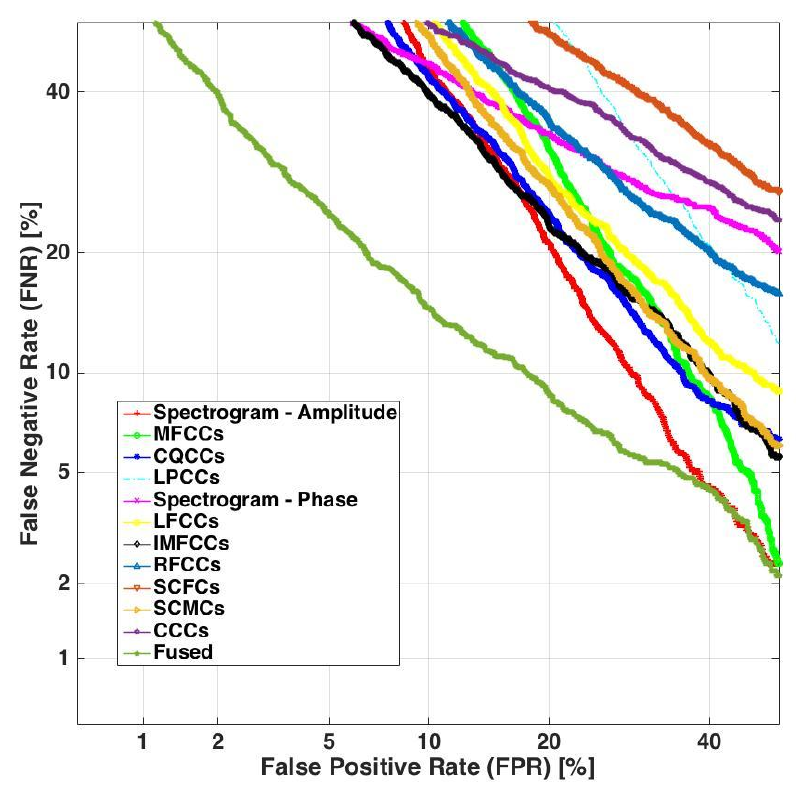}
\caption{DET Curves showing the performance of the spoofing detection models for Experiment~2.}
\label{Figure9}
\end{figure}

Based on the increased popularity of latent neural network representations, we expected a steep increase in performance when building the model using machine learned features, and including data augmentation. According to our results listed in Table~\ref{Table3} and Figure~\ref{Figure9}, this is, however, not the case for spoofing detection, at least not when using only these new, learned features. None of the models built using the new feature sets and augmented dataset, are able to match the performance of CQCC in Experiment~1. The performance of the model with an autoencoder based on CQCC features in Experiment~2 is slightly below that of Experiment~1. Models with average performance in Experiment~1 (e.g., SCFC, CCCs, LPCCs), keep a similar ranking in experiment 2, however, their absolute EER performance is compromised. 
The reason for this decreased performance may be related to the fact that there is a higher variance in the machine learned features compared to the original features, which can cause a drop in performance. This warrants further investigation.
The models based on spectral amplitude and MFCCs are the only ones that increase their performance in terms of EER when compared to that of Experiment~1.
 
 \begin{table}[h!]
\centering
\caption{Results from Experiment 2.}
\label{Table3}
\begin{tabular}{lc}
\toprule
\textbf{Features} & \textbf{EER }\\
\toprule
MFCCs & 24.2\\
CQCCs & \textbf{21.5}\\
Spectrogram - Amplitude &  \textbf{20.2}\\
Spectrogram - Phase & 26.5\\
LPCCs & 31.2\\
RFCCs & 27.9\\
LFCCs & 24.0\\
IMFCCs & 21.6\\
SCMCs & 23.0\\
SCFCs & 35.9\\
CCCs & 32.0\\
Fused result & \textbf{12.6}\\
\bottomrule
\end{tabular}

\vspace{6pt}
\footnotesize \emph{The features resulting in the three best performing\\ models are indicated in bold.}
\end{table}

\begin{figure*}[h!]
\centering
\includegraphics[scale=0.74]{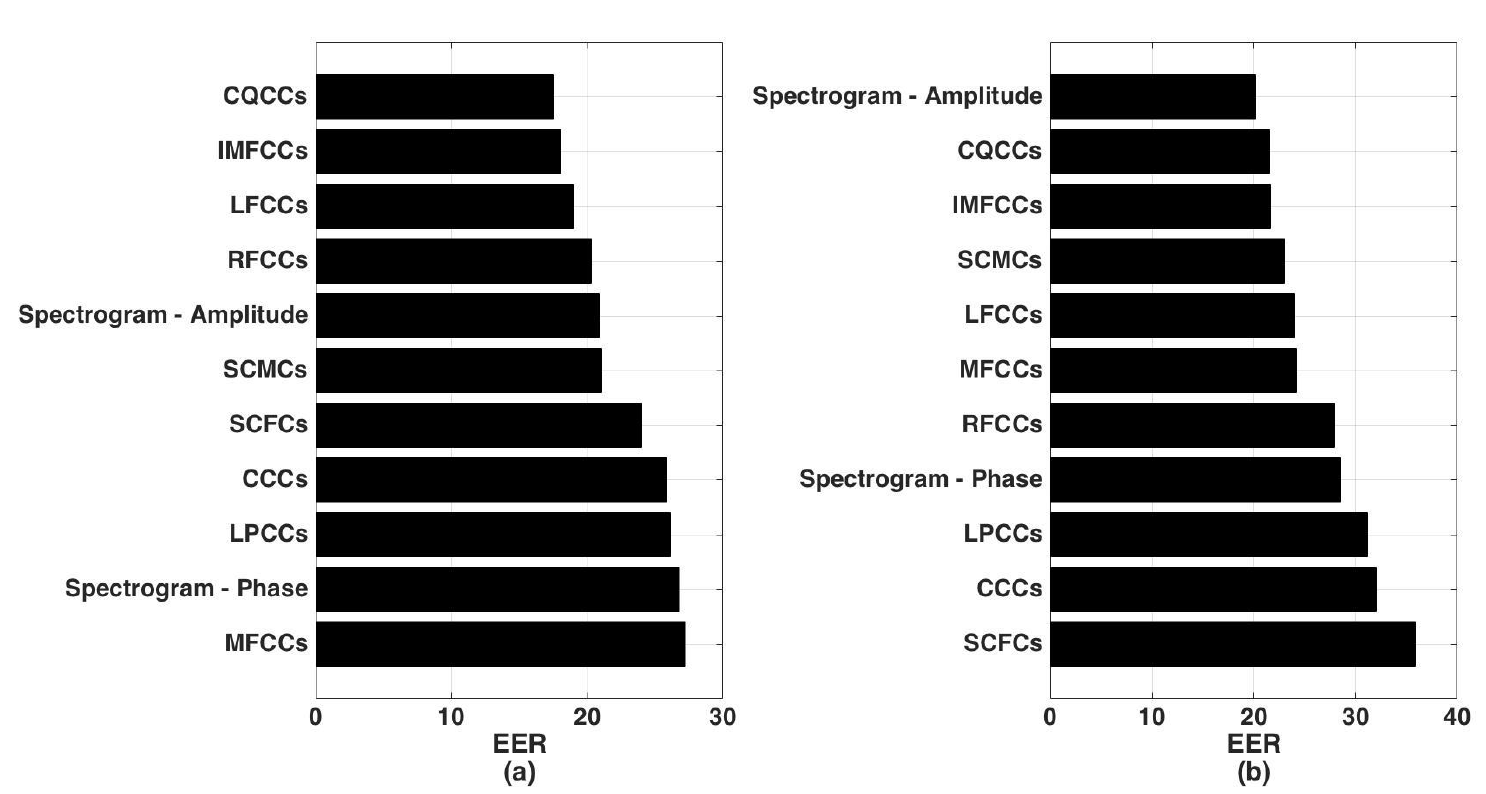}
\caption{Ranking of models built on different sets of audio features based on their EER performance: (a) with known features (Experiment 1) (b) with autoencoder features and data augmentation (Experiment 2).}
\label{Figure10}
\end{figure*}

When looking at the performance \emph{ranking} of models based on the different feature sets (see Figure~\ref{Figure10}), we see that it is different from the ranking in Experiment~1. For example, models based on spectral amplitude are now found to be the best performing models, whereas they are only the fifth best model in Experiment~1.


\subsection{Experiment 3: Performance of hybrid system}

\begin{figure}[h!]
\centering
\includegraphics[scale=0.8]{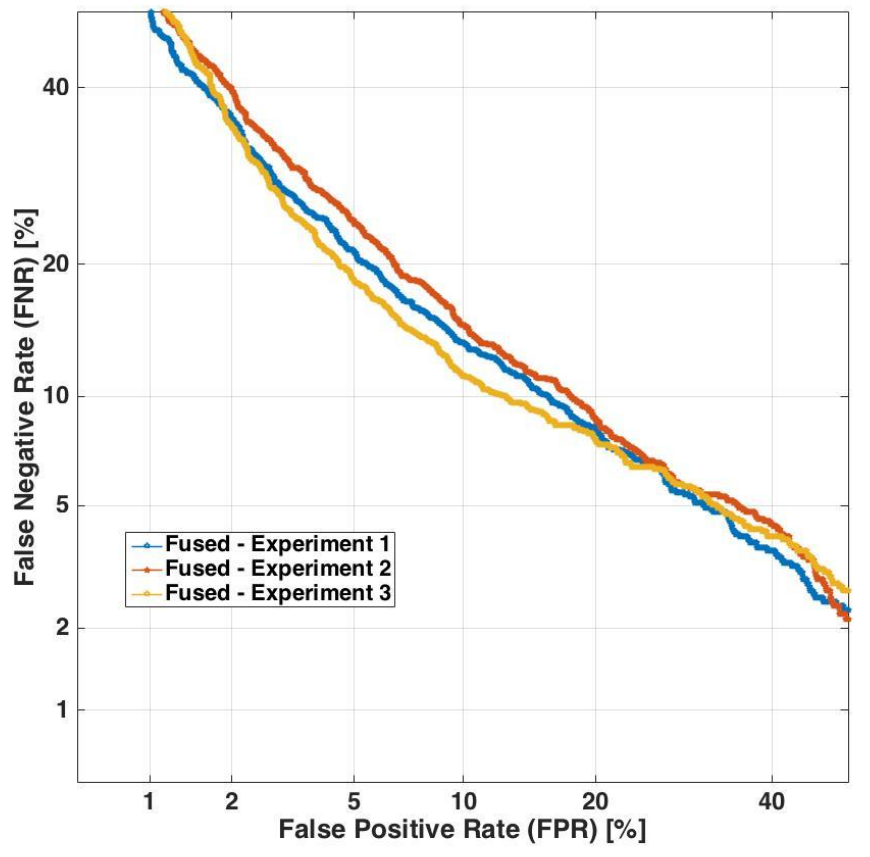}
\caption{DET Curves showing the performance of spoofing the detection models for Experiment~3.}
\label{Figure11}
\end{figure}

We explore the effectiveness of a hybrid system that combines both the predictions of the models built on known and machine learned features. The results of this experiment are shown in Table \ref{Table4} and the resulting DET curves are shown in Figure \ref{Figure11}. A superior performance is found when using a logistic regression fusion to combining the output of individual models. This could be attributed to the larger, more diverse feature space (coming from multiple models). The hybrid system outperforms all individual models, including the fused results from both Experiment 1 and 2. This confirms that each of the audio feature sets capture relevant aspects of audio signals that, when put together, form the most powerful model. 

\begin{table}[h!]
\centering
\caption{Results from Experiment 3.}
\label{Table4}
\begin{tabular}{cc}
\toprule
\textbf{Experiment} & \textbf{EER (Calibrated)}\\
\toprule
Experiment 1 (Fused) & 12.2\\
Experiment 2 (Fused) & 12.6\\
Experiment 1 and 2 (Fused) & 10.8\\
\bottomrule
\end{tabular}
\end{table}

\subsection{Comparison with state-of-the-art systems}
Table~\ref{Table5} compares the performance of our proposed system with other systems trained on the ASVSpoof2017 dataset in terms of EER. When comparing to other models based on GMM, such as System 3 \cite{table5at3}, we achieve better performance due to our unique feature set and dataset augmentation. For instance, with the traditional feature set from Experiment 1, a slightly better performance than System 3 is achieved, which uses the same core model, but less features. In Experiment 3, we further improve the accuracy of the proposed system by integrating an autoencoder. By using this to augment the training set and by fusing the results based on both traditional and learned features, the EER improves by 20\%, thus reaching 10.8. The only system that outperforms our proposed architecture integrates, in addition to GMM, a recursive neural networks \cite{table5at1}. It would be conceivable that the performance of said model would also improve when incorporating our feature learning and dataset augmentation method. This is a topic for future research. 

\begin{table*}[t!]
\centering
\footnotesize
\setlength\tabcolsep{4pt}
\caption{EER values obtained for different models on the ASVSpoof2017 dataset with various features.}
\label{Table5}
\begin{tabular}{lllllll}
\toprule
Systems & \begin{tabular}[c]{@{}l@{}} EER\end{tabular}  & Features & classifier & Fusion & Training Dataset \\
\toprule
\textbf{Proposed System} & \textbf{10.8} & \begin{tabular}[c]{@{}l@{}}\textbf{MFCCs, CQCCs,} \textbf{Spectrogram},\\\textbf{LPCCs, RFCCs, LFCCs, IMFCCs,} \\\textbf{SCMCs, SCFCs, CCCs and} \\\textbf{Autoencoder reconstructed features}\end{tabular} & \textbf{GMM}  & \textbf{Yes} & \textbf{Training}\\
System 1 \cite{table5at1} & 6.73  & Power Spectrum, LPCCs & CNN, GMM, RNN, Total Variation & Yes & Training \\
System 2 \cite{table5at2}& 12.34  & CQCCs, MFCCs, PLP & \begin{tabular}[c]{@{}l@{}}GMM-UBM, GSV-SVM, ivec-PLDA, \\ GBDT, Random Forest\end{tabular} & Yes & Training \\
System 3 \cite{table5at3} & 14.03  & \begin{tabular}[c]{@{}l@{}}MFCCs, IMFCCs, RFCCs, LFCCs,\\ PLPCCs, CQCCs, SCMCs, SSFCs\end{tabular} & GMM, FF-ANN & Yes & Training+Development \\
System 4 \cite{table5at4} & 14.66  & 

\begin{tabular}[c]{@{}l@{}}RFCCs, MFCCs, IMFCCs, LFCCs, \\ SSFCs, SCMCs\end{tabular} & GMM & Yes & Training+Development \\
System 5 \cite{table5at5}  & 15.97 &  Linear Filterbank Feature & \begin{tabular}[c]{@{}l@{}}GMM, CT-DNN with convolutional layer \\ \& time-delay layers\end{tabular} & Yes & Training \\
\begin{tabular}[c]{@{}l@{}}ASVSpoof Baseline\\ (B01)  \cite{11}\end{tabular} & 24.77 & CQCCs & GMM & No & Training+Development \\
\begin{tabular}[c]{@{}l@{}}ASVSpoof Baseline\\ (B02) \cite{11}\end{tabular} & 30.6 &  CQCCs & GMM & No & Training \\
\bottomrule
\end{tabular}
\end{table*}

\section{Conclusions}
\label{sec:conclusion}

We examine the effect of a large variety of audio features on the performance of a GMM-UBM based replay audio spoofing detection system. One of the goals of this paper is to pinpoint the most important features when building a \emph{robust} model that works on an `in the wild dataset', i.e., without having any information about the used replay spoofing technique. In addition to thoroughly comparing the models built on the different features, we also provide a clear procedure for proper pre-and postprocessing of these features, which we hope will be valuable to other researchers.

In our experiments, we explore both known audio features, and those learned by an autoencoder (i.e., using a feed forward neural network). The former includes some de-facto features often used this field, as well as some potentially new features that would be able to distinguish genuine speech from spoofed one. The included features are MFCCs, spectrogram, CQCCs, LPCCs, IMFCCs, RFCCs, LFCCs, SCFCs, SCMCs, and CCCs. Secondly, a novel representation for each of these feature sets is learned by an autoencoder. In addition, this autoencoder is used to augment the training set. 

The performance of each of the models built with these different feature sets is reported in terms of EER. When using only known features, or only autoencoder features, the resulting performance is around $12$ in terms of EER. When creating a hybrid system that incorporates both types of features, we achieve a superior performance of $10.8$. This competes with the current state-of-the-art and reiterates the importance of integrating different types of audio features, both known and machine learned, in order to develop a robust model for replay spoofing detection. 

\FloatBarrier



\ifCLASSOPTIONcaptionsoff
  \newpage
\fi

\bibliographystyle{IEEEtran}
\bibliography{paper}

\begin{IEEEbiography}[{\includegraphics[width=1in,height=1.25in,clip,keepaspectratio]{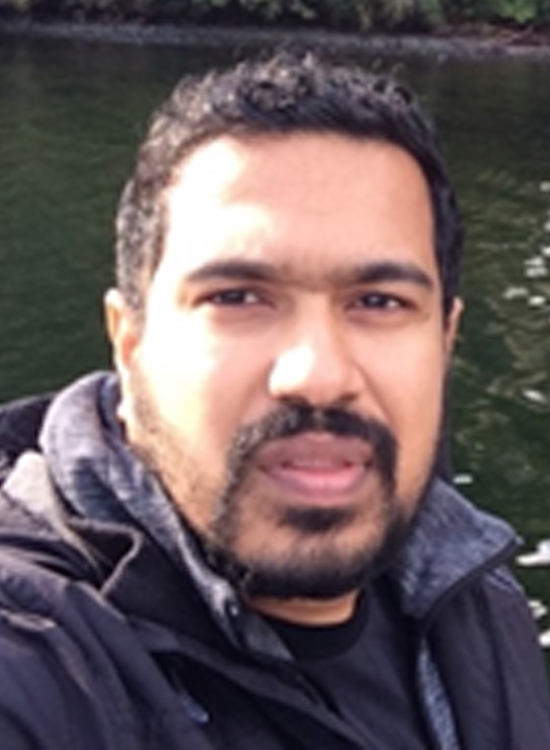}}]{Balamurali B T} is a postdoctoral research fellow working at the Singapore University of Technology and Design. He received his Ph.D. in Electrical and Computer Engineering from the university of Auckland, New Zealand in 2015. After his Ph.D, he worked as a researcher in Gastro intestinal group in Auckland Bio-engineering Institute and was a lecturer in Auckland university of technology. Prior to his Ph.D endeavor, he was a design and development engineer in Tata Elxsi, India. He is currently passionate about artificial intelligence and trying to solve a variety of problems related to bio-signal processing detection and classification, automatic speech/speaker recognition, spoofed-speech detection, blacklisted speaker identification, blind source separation, music classification, fluid flow classification, fruit ripeness detection etc. \end{IEEEbiography}

\begin{IEEEbiography}[{\includegraphics[width=1in,height=1.25in,clip,keepaspectratio]{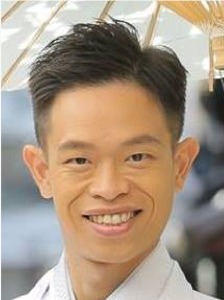}}]{Kin Wah Edward LIN} is now an AIST Postdoctoral Researcher, working at the National Institute of Advanced Industrial Science and Technology (AIST), Japan.  He received his Ph.D. degree from Singapore University of Technology and Design (SUTD) in 2018. Edward has obtained many scholarships to study in top schools and universities in Singapore and Hong Kong. He also has 3 years undergraduate teaching experience and 3 years working experience in Hong Kong IT Industry. He has published 2 WiFi-related papers during his MPhil study and 8 Audio-related paper during his PhD study. His research interests now include Audio-related iOS app and Singing Voice Separation.
\end{IEEEbiography}

\begin{IEEEbiography}[{\includegraphics[width=1in,height=1.25in,clip,keepaspectratio]{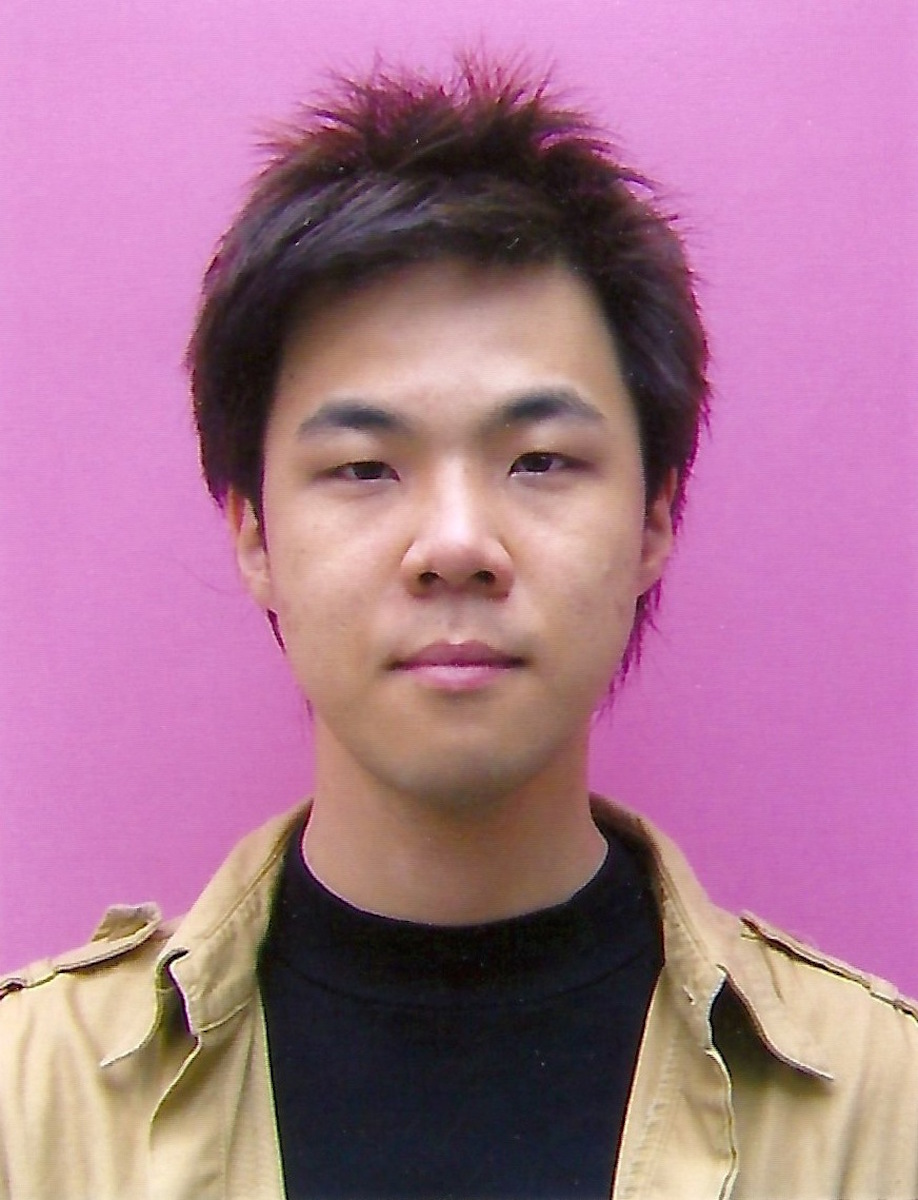}}]{Simon Lui}  received his Ph.D. degree in Computer Science from The Hong Kong University of Science and Technology (HKUST) in 2011. He is currently the Director of Tencent Music Entertainment Group (TME), and Adjunct Assistant Professor of the Singapore University of Technology and Design (SUTD). Simon was an Assistant Professor of SUTD in 2012-2018, and a visiting scholar of the Massachusetts Institute of Technology (MIT) CSAIL in 2012-2013. His primary research interests are machine learning and music information retrieval. Simon developed several best selling apps in the iOS stores in 7 countries. His business story was reported by CNN International and featured in IEEE Institute. 
\end{IEEEbiography}
\vfill\null

\begin{IEEEbiography}[{\includegraphics[width=1in,height=1.25in,clip,keepaspectratio]{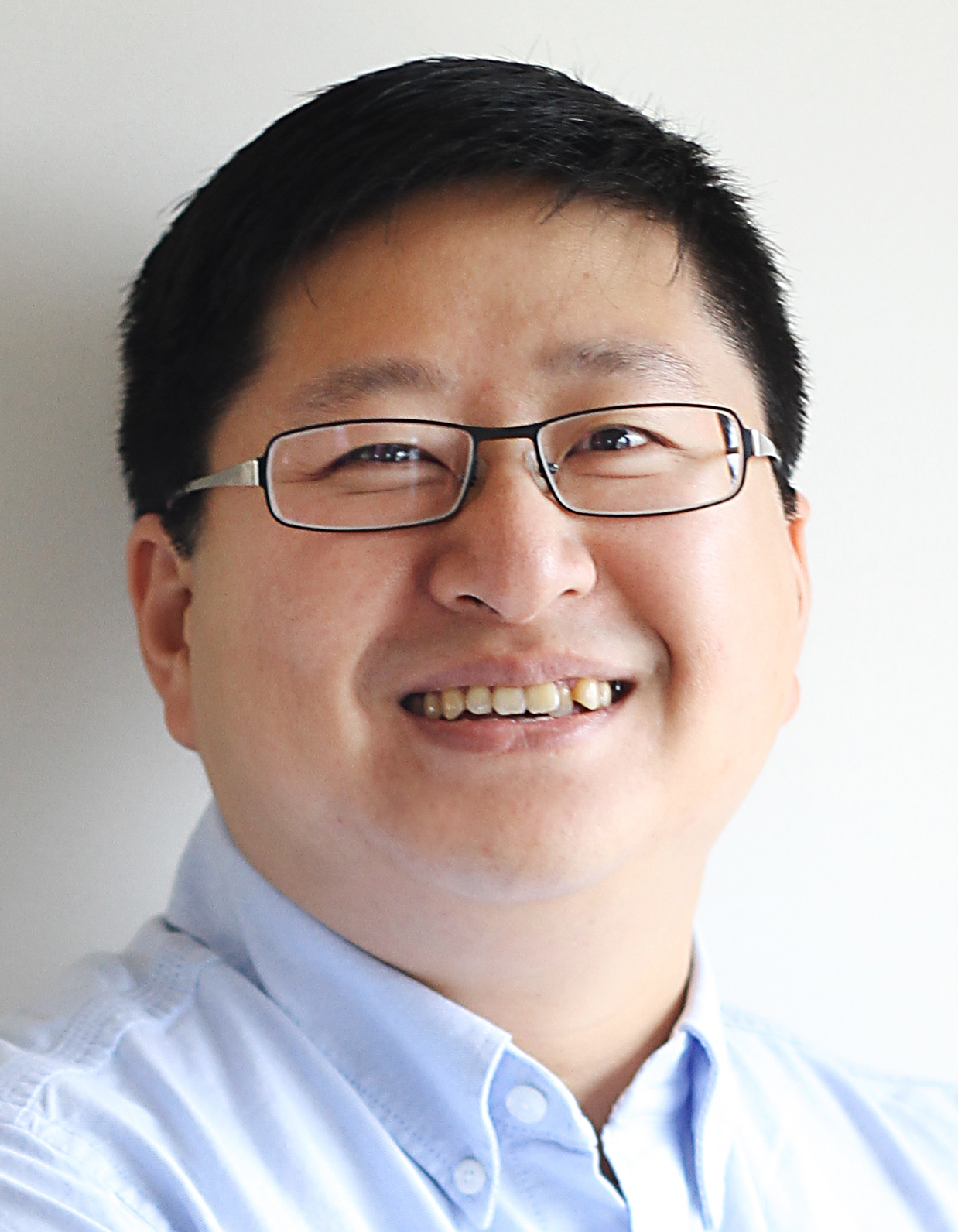}}]{Jer-Ming Chen} is an Assistant Professor at the Singapore University of Technology and Design. He received his Ph.D. in Applied Physics from the University of New South Wales in 2010. His primary research interest is in the interaction of coupled resonators in acoustics. Jer-Ming has also written lay-language scientific papers and has featured in the international media and popular press, including newspapers (e.g. New York Times, UK Telegraph, Sydney Morning Herald), TV and radio documentaries (BBC, ABC, Network Ten), and popular science magazines (Physics Today, Scientific American, The Straight Dope).
\end{IEEEbiography}

\begin{IEEEbiography}[{\includegraphics[width=1in,height=1.25in,clip,keepaspectratio]{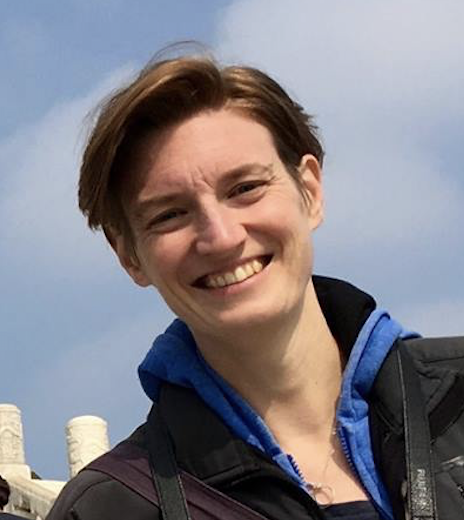}}]{Dorien Herremans} (SM'17) is an Assistant Professor at Singapore University of Technology and Design, with a joint appointment at the Institute of High Performance Computing at the Agency for Science Technology and Research, A*STAR. In 2015, she was awarded the individual Marie-Curie Fellowship for Experienced Researchers, and worked at the Centre for Digital Music, Queen Mary University of London. Prof. Herremans received her PhD in Applied Economics from the University of Antwerp. After graduating as a commercial engineer in management information systems at the University of Antwerp in 2005, she worked as a Drupal consultant and was an IT lecturer at Les Roches University in Bluche, Switzerland. Prof. Herremans' research focuses on the intersection of machine learning/optimization and digital music/audio. She is a Senior Member of the IEEE and co-organizer of the First International
Workshop on Deep Learning and Music as part of IJCNN, as well as guest editor for Springer's Neural Computing and Applications.
\end{IEEEbiography}
\vfill\null
\EOD

\end{document}